\documentclass[aps,pra,reprint]{revtex4-2}
\usepackage[colorlinks,linkcolor=blue,citecolor=blue]{hyperref}
\usepackage[T1]{fontenc}
\usepackage{amsmath,bm}
\usepackage{balance} 
\usepackage{graphicx}
\usepackage{amssymb}

\newcommand{\na}{\bm\nabla}
\renewcommand{\Im}{{\rm Im\,}} 
\renewcommand{\Re}{{\rm Re\,}}

\begin{document}
\title{Dual Transverse Vortex-Spin-Momentum Locking in Rotating-Wave Structured Light}
\author{Zhen-Lai Wang}
\email{wangzhenlai@hbpu.edu.cn}
\affiliation{Center for Fundamental Physics and School of Artificial intelligence, Hubei Polytechnic University, Huangshi 435003, China}
\date{\today}

\begin{abstract}
We theoretically investigate the transverse Poynting momentum and transverse spin angular momentum in circularly polarized rotating-wave structured light, and reveal a pair of \emph{dual} transverse vortex-spin-momentum locking. The transverse Poynting momentum exhibits a $\mathbb{Z}_2 \times \mathbb{Z}_2$ symmetric locking structure, being controlled independently by either the polarization or vortex handedness, yet invariant under their simultaneous reversal. Meanwhile, the transverse spin angular momentum is independent of the polarization but endowed with a crucial vortex-dependence inheriting such vortex-dependence from the Poynting momentum via curl operation. This finding expands the understanding of transverse optical linear and angular momenta in structured light and opens a new paradigm for vortex-spin-controlled optical manipulation.
\end{abstract}

\maketitle

\section{Introduction}\label{I}

In free-space optics, linear and angular momenta are typically locked to the propagation direction. For a plane wave or a paraxial beam, the dominant momentum is directed forward; spin angular momentum (SAM), associated with circular polarization, and intrinsic orbital angular momentum (OAM), associated with vortex phase, are both aligned with the propagation direction. In particular, SAM/OAM is either parallel or antiparallel to the propagation direction, depending on the polarization/vortex handedness~\cite{ Allen}. 

However, this propagation-oriented conventional picture fails for structured optical fields, where linear and angular momenta can acquire transverse components orthogonal to the propagation direction. Two anomalies are of particular interest.

One is the polarization-dependent transverse momentum, suggested for evanescent waves~\cite{Blio14} and other structured fields~\cite{Beksh15,Saha16,Wang18,Pal20}. It can be regarded as the spin momentum, a part of the Poynting momentum, introduced by Belinfante~\cite{Beli40} to symmetrize the canonical energy-momentum tensor. As a remarkable conclusion~\cite{Berr09}, the spin momentum is fully determined by the curl of SAM. Hence, the transverse spin momentum inherits its polarization dependence from SAM structure, and can reverse with polarization handedness. The spin momentum has long been misunderstood as virtual~\cite{Beli40,Berr09,Ohani86,BBN13}, on account of it carrying no energy flow, exerting no force a dipole particle, or generating anything beyond SAM. However, such elusive transverse spin momentum manifests as a lateral force, as verified and detected experimentally by helicity-dependent force measurements on a nanocantilever~\cite{Anto16} and on a Mie particle~\cite{Liu18}.

The other is the helicity-independent transverse SAM~\cite{Blion15}, emerging in surface modes~\cite{Blio12,Kim12,B15sci,Shi21}, evanescent waves~\cite{Blio14,Meche16}, and other structured fields~\cite{Beksh15,Saha16,Wang18,Pal20,Neuge15,Aiel15,Abuj20,ShiD23,ShiL23,Chen23}. In turn, such transverse SAM is uncovered to be associated with the curl of the Poynting momentum~\cite{Shi21,Beksh22,Verno24,Xiong25}. It directly drives the emergence of  the ``spin-momentum locking'' in evanescent-field modes~\cite{B15sci,Shi21,Aiel15,Meche16}, which enables spin-controlled unidirectional excitation of surface and guided modes~\cite{B15sci,Rodr13,Neuge16}.

Beyond transverse SAM, structured optical fields can also carry vortex-dependent transverse OAM~\cite{Vo26,Wang24}. As noted above, both transverse SAM and transverse momentum arise from the inhomogeneity of the field intensity rather than directly from the vortex phase. Given this clear distinction in physical origin, the question naturally arises is whether an optical vortex can control transverse SAM and transverse momentum. This question is both of theoretical and practical interest, with applications ranging from spin-controlled directional excitation to chirality-selective light-matter interactions.

In this work, we propose a pair of \emph{dual} transverse vortex-spin-momentum locking in circularly polarized rotating-wave structured light~\cite{Wang24,Ceper92}. Unlike the conventional spin-momentum locking in evanescent-field modes, such effect reveals that an optical vortex can control both the transverse SAM and the transverse momentum. In particular, the transverse Poynting momentum (or Belinfante spin momentum) exhibits a $\mathbb{Z}_2 \times \mathbb{Z}_2$ symmetric locking structure: it can be controlled separately by either the polarization or vortex handedness, while remaining invariant under their simultaneous reversal. On the other hand, the sign of the transverse SAM varies with the vortex handedness but not with the polarization of rotating-wave structured light. This crucial vortex-dependence of transverse SAM can be understood as inheriting from the Poynting momentum structure via curl operation, and so leads to vortex-spin-momentum locking as well. Taken together, circularly polarized rotating-wave structured light thus manifests a pair of \emph{dual} transverse vortex-spin-momentum locking in free space. This finding expands our understanding of transverse momentum and transverse SAM in structured light and opens a new paradigm for vortex-spin-controlled optical manipulation~\cite{Wang25,Mur26}.

\section{Circularly-polarized Rotating-wave structured light}\label{II}

\begin{figure}[ht]
    \centering
    \includegraphics[width = 0.49 \columnwidth]
    {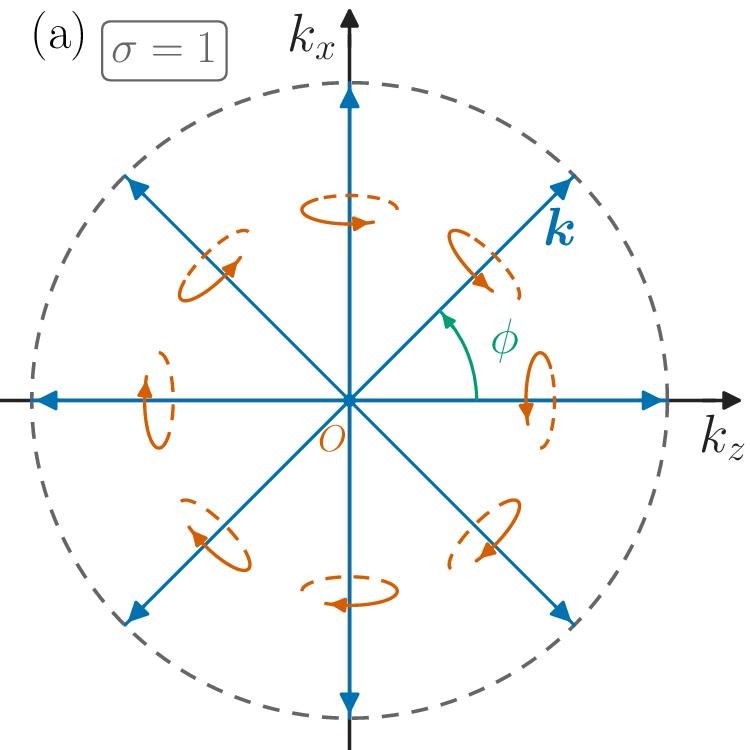}   \hfill\hfill
    \includegraphics[width = 0.49 \columnwidth]
    {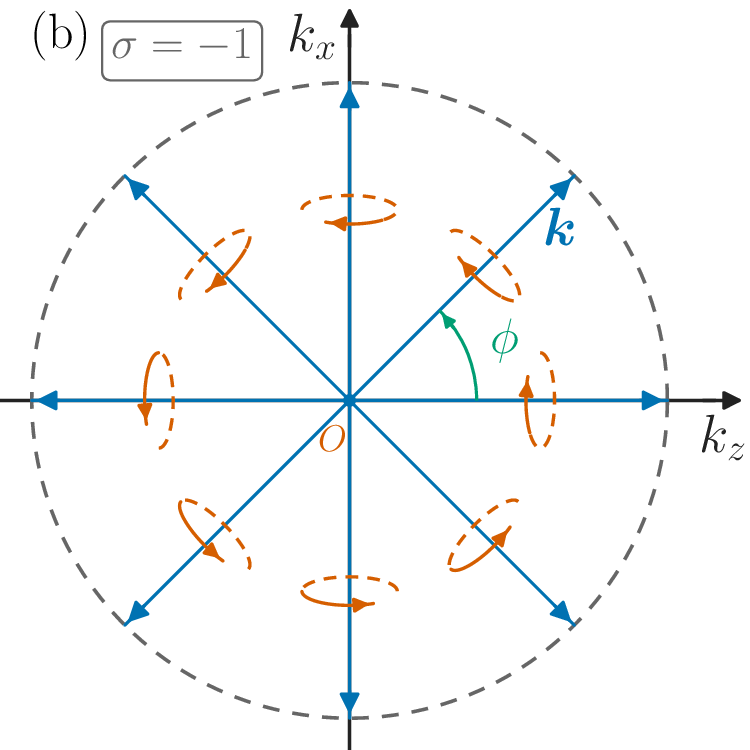}     
    \caption{Schematic diagrams of the two basic modes of circular polarization. For each plane-wave component, the electric and magnetic fields rotate about the wave vector $\bm{k}$, viewed along $\bm{k}$: (a) Right-handed mode, clockwise; (b) Left-handed mode, counterclockwise.}
    \label{CPL}
\end{figure}

We begin with a monochromatic rotating-wave structured light beam propagating along the azimuthal direction in the $(z,x)$ plane. It is formed by superposing plane waves with wave vectors $\bm{k}$ distributed on a circle in momentum space, with the corresponding azimuthal phase difference $\ell\phi$, where $\ell=0,\pm 1,\pm 2,...$ is the topological charge and $\phi$ is the azimuthal angle in $\bm k$ space. For a monochromatic scalar rotating wave, the complex wave function can be written as
\begin{align}
\psi_{\ell}(\bm{r},t) &= \int_{0}^{2\pi} e^{ik\cos\phi z + ik\sin\phi x +i\ell\phi-i\omega t} {\rm d}\phi\nonumber\\&=2\pi{i}^{\ell}J_{\ell}(k\rho)e^{i\ell\varphi-i\omega t}.\label{eq:rw}
\end{align}
Here $J_{\ell}(k\rho)$ is the $\ell$th-order Bessel function,  $k=\omega/c$ is the wave number with $\omega$ the frequency and $c$ the phase velocity, and $(\rho,\varphi)$ are the polar coordinates in the $(z,x)$ plane with $z=\rho\cos\varphi$ and $x=\rho\sin\varphi$. The sign of $\ell$ determines the azimuthal direction of propagation: $\ell>0$ and $\ell<0$ correspond to propagation along $+\varphi$ and $-\varphi$, respectively.

Now we construct the vector electromagnetic fields of rotating wave. For a linearly polarized rotating-wave structured light beam, the two supported dual modes~\cite{Wang24} are: (i) the TM mode, which has a $y$-directed magnetic field and electric field confined to the $(z,x)$ plane; and (ii) its dual TE mode, with a $y$-directed electric field and magnetic field in the $(z,x)$ plane. For later reference, the relevant expressions of such two dual modes are restated explicitly here:
\begin{align}
\bm{B}_{\ell}^{\rm TM}&=\psi_{\ell}(\bm{r},t)\hat{\bm{y}},\qquad\bm{E}_{\ell}^{\rm TM}=\frac{i}{k}\na\times\bm{B}_{\ell}^{\rm TM},\label{eq:tm}\\
\bm{E}_{\ell}^{\rm TE}&=-\psi_{\ell}(\bm{r},t)\hat{\bm{y}},\quad~\bm{B}_{\ell}^{\rm TE}=-\frac{i}{k}\na\times\bm{E}_{\ell}^{\rm TE}.\label{eq:te}
\end{align}
Here, $\hat{\bm{y}}$ is the unit vector in the $y$ direction.

To incorporate spin (circular polarization) degrees of freedom, we introduce a helicity parameter $\sigma = \pm 1$ representing the two circular polarization states. Parameterized by the single helicity index $\sigma=\pm 1$, the circularly polarized rotating-wave beams can be formed by superposing the two linearly polarized modes above with equal amplitude and a $\pi/2$ phase difference:
\begin{equation}
\{\bm{E}^{(\sigma,\ell)},\bm{B}^{(\sigma,\ell)}\}=\frac{\{\bm{E}_{\ell}^{\rm TM},\bm{B}_{\ell}^{\rm TM}\}-i\sigma \{\bm{E}_{\ell}^{\rm TE},\bm{B}_{\ell}^{\rm TE}\}}{\sqrt{1+\sigma^2}}.\label{eq:spineb}
\end{equation}
This construction guarantees that for $\sigma=+1$ ($\sigma=-1$), each plane-wave spectrum of circularly polarized rotating-wave beam corresponds to right-handed (left-handed) circular polarization in a well-defined sense; see Figs.~\ref{CPL}(a) and \ref{CPL}(b). In particular, if $\sigma = 0$, the linearly polarized case is recovered, and so can be taken as a reference for our analysis.

\section{vortex-spin locking of transverse Poynting momentum}\label{III}

We now recall the two fundamental momentum densities of light in free space: the Poynting and canonical momenta. For a monochromatic optical field in free space, the time-averaged Poynting momentum density reads 
\begin{equation}
\bm{\mathcal{P}}=gk\Re\big[{\bm E}^{*}\times {\bm B}\big].\label{eq:poy}
\end{equation}
Hereafter, we use Gaussian units with $g=(8\pi\omega)^{-1}$. Conversely, the canonical momentum density is defined as
\begin{equation}
\bm{P}=\frac{g}{2}\Im\big(E^{*}_{j} \na E_{j}+B^{*}_{j}\na B_{j}\big),\label{eq:can}
\end{equation}
which naturally yields the well-known spin-orbit decomposition of the total angular momentum density $\bm{\mathcal{J}}= \bm{L} +\bm{S}$, with the OAM part
\begin{equation}
\bm{L}=\frac{g}{2}\Im\big[E^{*}_{j} (\bm{r}\times\na) E_{j}+B^{*}_{j}(\bm{r}\times\na)B_{j}\big]\label{eq:oam}.
\end{equation}
and the spin part
\begin{equation}
\bm{S}=\frac{g}{2}\Im\big[\bm{E}^{*}\times\bm{E}+\bm{B}^{*}\times\bm{B}\big]\label{eq:sam}.
\end{equation}
Here the index $j$ runs over the three spatial coordinates. 

Such two familiar momentum densities are linked by
\begin{equation}
\bm{\mathcal{P}}=\bm{P}+ \bm{P}_{\rm s}=\bm{P}+\frac{1}{2}\na\times\bm{S}\label{eq:pps},
\end{equation}
in which the Poynting momentum density is separated into orbital (canonical) and spin contributions with strikingly different physical behavior. The canonical momentum density, tied to the field phase gradients and  the local wavevector~\cite{Berr09}, responsible for radiation pressure and momentum transfer~\cite{Gordo73,Ashk83,Chau00,Beksha12}. The spin momentum density, by contrast, is a curl-type vector and thus inherently divergence-free, and consequently yields no net energy flow. 

Substituting Eq.~\eqref{eq:spineb} and Eqs.~(\ref{eq:rw})--(\ref{eq:te}) into Eqs.~(\ref{eq:poy}) and~(\ref{eq:can}), and removing the common factor $g\pi^2$, one can obtain the densities of the Poynting and canonical momenta in circularly polarized rotating-wave beam:
\begin{align}
\bm{\mathcal{P}}^{(\sigma,\ell)}&=k\sigma(J_{\ell-1}^2-J_{\ell+1}^2)\hat{\bm{y}} + \frac{4\ell}{\rho}J_{\ell}^2\hat{\bm{\varphi}},\label{eq:poyl}\\
\bm{P}^{(\sigma,\ell)}&=\frac{1}{\rho}\big[(\ell-1)J_{\ell-1}^2+2\ell J_{\ell}^{2}+(\ell+1)J_{\ell+1}^2\big]\hat{\bm{\varphi}}\label{eq:pcl}
\end{align}
Henceforth, the explicit $k\rho$ argument in the Bessel functions is dropped for ease of notation and $\hat{\bm{\varphi}}$ is the unit vector in azimuthal direction. Interestingly, the canonical momentum density~\eqref{eq:pcl} is independent of the helicity index and purely oriented along the azimuthal direction within the $(z, x)$ plane. However, the Poynting momentum density acquires an extra $y$-directed transverse momentum. Yet this transverse momentum occurs even though the light beam itself propagates only in the $(x,z)$ plane. Remarkably, such transverse Poynting momentum depends on the helicity index and stems entirely from the spin momentum, since the canonical momentum carries no $y$-component, via Eq.~\eqref{eq:pps}. 

Let us now revisit the $\sigma$-dependent transverse component of the Poynting momentum:
\begin{align}
\mathcal{P}_y^{(\sigma,\ell)} = k\sigma(J_{\ell-1}^2-J_{\ell+1}^2)= \frac{2\sigma \ell}{k\rho}\frac{{\rm d}J_{\ell}^2}{{\rm d}\rho},
\label{eq:Py}
\end{align}
which constitutes one of the central results of this paper. It demonstrates three distinct locking properties.

\emph{First}, the transverse Poynting momentum is \emph{strictly locked to the polarization handedness} at every local point within the $(z, x)$ plane, with opposite polarization handedness states producing opposite transverse momentum:
\begin{equation}
\mathcal{P}_y^{(-,\ell)} = -\,\mathcal{P}_y^{(+,\ell)},
\label{eq:spinlock}
\end{equation}

\emph{Second}, the transverse Poynting momentum depends on the topological charge $\ell$ through the difference $J_{\ell+1}^2-J_{\ell-1}^2$. Since $\ell$ determines the azimuthal direction of propagation in the rotating-wave beam, reversing $\ell$ reverses the sign of $\mathcal{P}_y$:
\begin{equation}
\mathcal{P}_y^{(\sigma,-\ell)} = -\,\mathcal{P}_y^{(\sigma,\ell)}.
\label{eq:llock}
\end{equation}
The sign reversal originates from the exchange of the indices $\ell-1$ and $\ell+1$ under $\ell \to -\ell$, which is a direct consequence of the Bessel reflection formula for integer-order Bessel functions, $J_{-n}^2 = J_n^2$.

\emph{Third}, combining Eqs.~(\ref{eq:spinlock}) and~(\ref{eq:llock}), the transverse Poynting momentum is governed by the \emph{joint chirality}
$\sigma \ell$:
\begin{equation}
\mathcal{P}_y^{(\sigma,\ell)} \;\propto\; \sigma\,\ell,
\label{eq:chirality}
\end{equation}
and so reversing both $\sigma$ and $\ell$ simultaneously leaves the transverse momentum unchanged:
\begin{equation}
\mathcal{P}_y^{(-\sigma,-\ell)} = \mathcal{P}_y^{(\sigma,\ell)}.
\label{eq:doubleflip}
\end{equation}

\begin{figure}[ht]
    \centering
    \includegraphics[width = 0.98 \columnwidth]
    {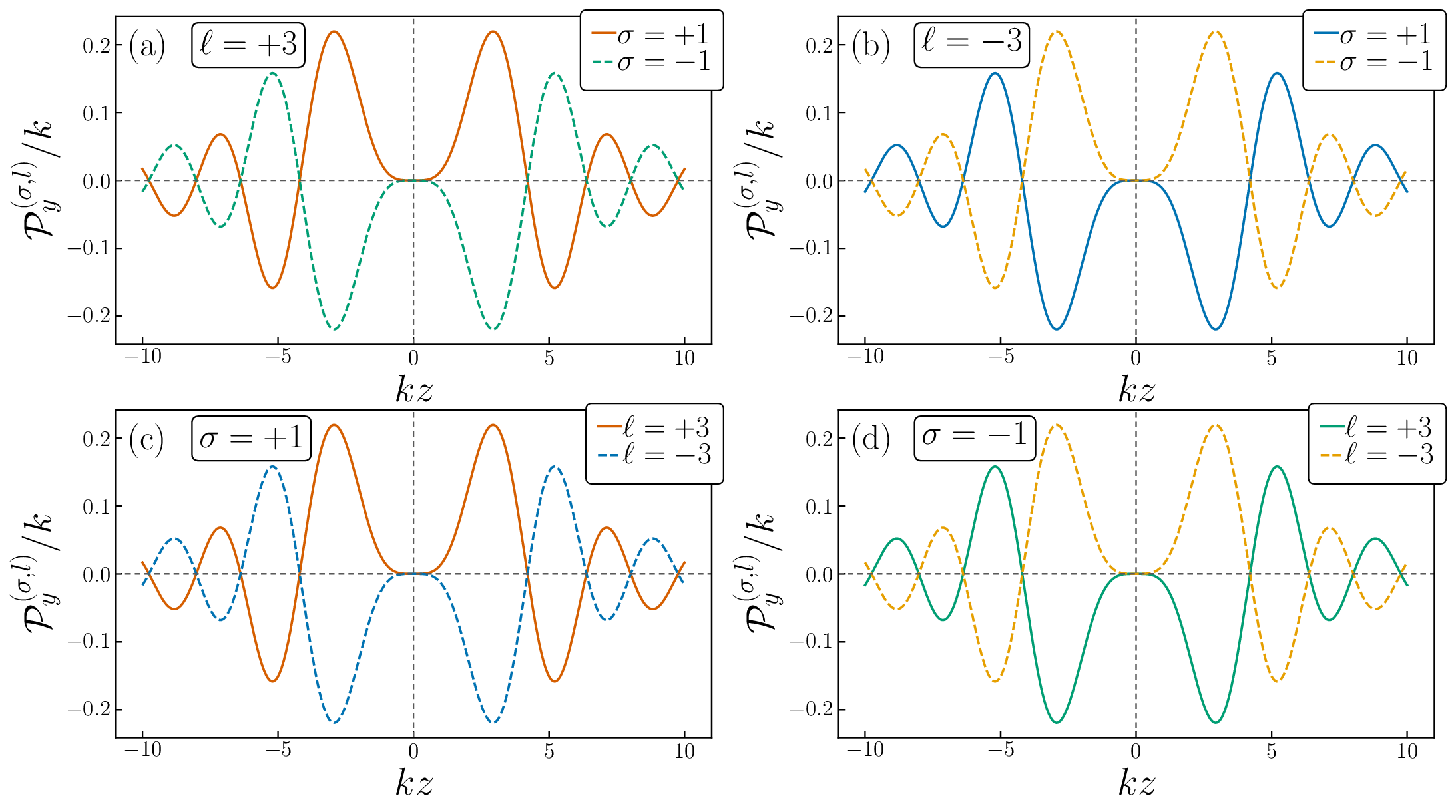}           
    \caption{The $\mathbb{Z}_2 \times \mathbb{Z}_2$ symmetric locking structure of transverse Poynting momentum.}
    \label{SV-Lock}
\end{figure}

As derived in Eqs.~(\ref{eq:spinlock})--(\ref{eq:doubleflip}), the system exhibits a $\mathbb{Z}_2 \times \mathbb{Z}_2$ symmetric locking structure: the transverse Poynting momentum can be controlled independently by the polarization or vortex handedness, and remains invariant under their simultaneous reversal, as illustrated in Figs.~\ref{SV-Lock}(a)--\ref{SV-Lock}(d). The product $\sigma \ell$ serves as the joint chirality index of the rotating-wave beam, conserved under simultaneous reversal of the polarization and vortex handedness, and uniquely determines the sign of the transverse Poynting momentum.

Despite its strong local dependence on $\sigma$ and $\ell$, the integral of $\mathcal{P}_y^{(\sigma,\ell)}$ over the entire $(z,x)$ plane vanishes exactly:
\begin{align}
\langle \mathcal{P}_{y}^{(\sigma,\ell)} \rangle &= \int_0^{2\pi} {\rm d}\varphi \int_0^\infty \mathcal{P}_{y}^{(\sigma,\ell)}\,\rho\,{\rm d}\rho \nonumber\\
&=k\sigma\int^{2\pi}_{0}{\rm d}\varphi\int^{\infty}_{0}(J_{\ell-1}^2-J_{\ell+1}^2)\rho{\rm d}\rho\nonumber\\
&= \frac{8\pi \ell \sigma}{k}\int^{\infty}_{0}J_{\ell}(k\rho){\rm d}J_{\ell}(k\rho)=0.
\label{eq:integral}
\end{align}
This integral vanishes identically, ensuring that the total transverse momentum is exactly zero for both polarization states and the topological charges. Although it is purely local and oscillatory, the transverse momentum $\mathcal{P}_y^{(\sigma,\ell)}$ is consistent with energy-momentum conservation in free space.

\section{Transverse SAM, OAM and spin-orbit interaction}\label{IV}

To confirm the spin-orbital-momentum relationship in the rotating-wave field, we now calculate the SAM and OAM densities. For completeness, the time-averaged energy density of a monochromatic optical field in free space is given by
\begin{equation}
 W=\frac{g\omega}{2}\big(|\bm{E}|^2+|\bm{H}|^2\big).\label{eq:web}\\
\end{equation}
Substituting the electromagnetic fields~(\ref{eq:spineb}) into the definitions of the energy~(\ref{eq:web}), SAM~(\ref{eq:sam}) and OAM~(\ref{eq:oam}) densities, and omitting the common factor $g\pi^2$, we have
\begin{equation}
W^{(\sigma,\ell)}=(J_{\ell-1}^2+2J_{\ell}^2+J_{\ell+1}^2)\omega,
\end{equation}
\begin{equation}
\bm{S}^{(\sigma,\ell)}=(J_{\ell-1}^2-J_{\ell+1}^2)\hat{\bm{y}} + \frac{4\ell\sigma}{k \rho}J_{\ell}^2\hat{\bm{\varphi}},\label{eq:sl}
\end{equation}
\begin{equation}
\bm{L}^{(\sigma,\ell)}=\big[(\ell-1)J_{\ell-1}^2+2\ell J_{\ell}^{2}+(\ell+1)J_{\ell+1}^2\big]\hat{\bm{y}} - yP_{\varphi}^{(\sigma,\ell)}\hat{\bm{\rho}}.
\label{eq:Ll}
\end{equation}
Here $\hat{\bm{\rho}}$ is the unit vector in the radial direction.

As shown in Eq.~\eqref{eq:sl}, the SAM density consists of two parts: the transverse $y$-component and the azimuthal component within the \((z,x)\) plane. On integration over the \((z,x)\) plane, the azimuthal contribution vanishes identically, owing to its uniform and purely circular distribution. The remaining transverse \(y\)-component represents the intrinsic transverse SAM, which is $\sigma$-independent and persists even for linearly polarized rotating-wave structured light. The OAM density in~\eqref{eq:Ll} also exhibits a dual structure: the transverse $y$-component and the radial component within the \((z,x)\) plane. The radial component is directly tied to the canonical momentum density, which also vanishes upon integration over the \((z,x)\) plane because its direction rotates uniformly with the azimuthal angle.

Consequently, only the transverse \(y\) components of the SAM and OAM densities can survive the integration over the entire $(z,x)$ plane. The integral SAM and OAM thus reduce to
\begin{equation}
\langle \bm{S}^{(\sigma,\ell)}\rangle=\langle S_{y}^{(\sigma,\ell)}\rangle\,\hat{\bm{y}}=\bm{0},
\end{equation}
\begin{equation}
\langle \bm{L}^{(\sigma,\ell)}\rangle=\langle L_{y}^{(\sigma,\ell)}\rangle\,\hat{\bm{y}}
=\frac{\ell}{\omega}\langle W^{(\sigma,\ell)}\rangle\,\hat{\bm{y}},
\end{equation}
where use has been made of \(\langle S_{y}^{(\sigma,\ell)}\rangle=0\), by analogy with the vanishing integral of $\mathcal{P}_y^{(\sigma,\ell)}$ obtained above. Therefore, in the \((z,x)\) plane only the transverse \(y\)-direction contributes to the integral angular momentum, and more importantly, the integral total transverse angular momentum belongs entirely to the intrinsic transverse OAM:
\begin{equation}
\langle \mathcal{J}_{y}^{(\sigma,\ell)}\rangle=\langle L_{y}^{(\sigma,\ell)}\rangle=\frac{\ell}{\omega}\langle W^{(\sigma,\ell)}\rangle. 
\end{equation}
This result is consistent with the local conservation law of transverse angular momentum from the sum of the transverse SAM~(\ref{eq:sl}) and OAM~(\ref{eq:Ll}) densities
\begin{equation}
\mathcal{J}_{y}^{(\sigma,\ell)}=L_{y}^{(\sigma,\ell)}+S_{y}^{(\sigma,\ell)}=\frac{\ell}{\omega}W^{(\sigma,\ell)}.
\end{equation}

Physically, this local conservation law signifies a transverse spin-orbit interaction~\cite{Bliok10,Fu24} in rotating-wave structured light, where the transverse SAM and OAM densities can interconvert with their energy-weighted sum conserved. Such a conversion occurs without any light-matter interaction and is therefore a purely free-space effect. While the transverse SAM and OAM densities can mutually convert at the local level, their integrated \(y\)-components are robust, with the spin contribution averaging to zero and the orbital contribution fixed by the topological charge \(\ell\). Thus, the rotating-wave structured light of topological charge $\ell$ carries an intrinsic transverse OAM of \(\ell\hbar\) per photon if quantizing the photon energy as \(\hbar\omega\).

\section{Dual transverse vortex-spin-momentum locking}\label{V}

Under the reversal of the topological charge $\ell \to -\ell$, we find 
\begin{equation}
\bm{S}^{(\sigma,-\ell)} = -\bm{S}^{(\sigma,\ell)},\label{eq:s-s}
\end{equation}
\begin{equation}
\bm{\mathcal{P}}^{(\sigma,-l)} = -\bm{\mathcal{P}}^{(\sigma,\ell)}\label{eq:p-p}
\end{equation}
Equations~(\ref{eq:s-s}) and~(\ref{eq:p-p}) show that both the SAM density and the Poynting momentum density reverse their signs when the vortex topological charge is reversed, while leaving spatial profiles unchanged. This sign reversal suggests the intrinsic nature of the SAM density in rotating-wave structured light originates from the vortex topology rather than from local polarization alone. Given that the vortex topological charge determines the OAM, the simultaneous sign reversal of the SAM and Poynting momentum densities provides direct evidence for the orbital-spin-momentum locking triad: the OAM, the SAM, and the Poynting momentum degrees of freedom are locked in such a way that reversing the vortex handedness reverses the entire triad.

\begin{figure}[ht]
    \centering
    \includegraphics[width = 0.98 \columnwidth]
    {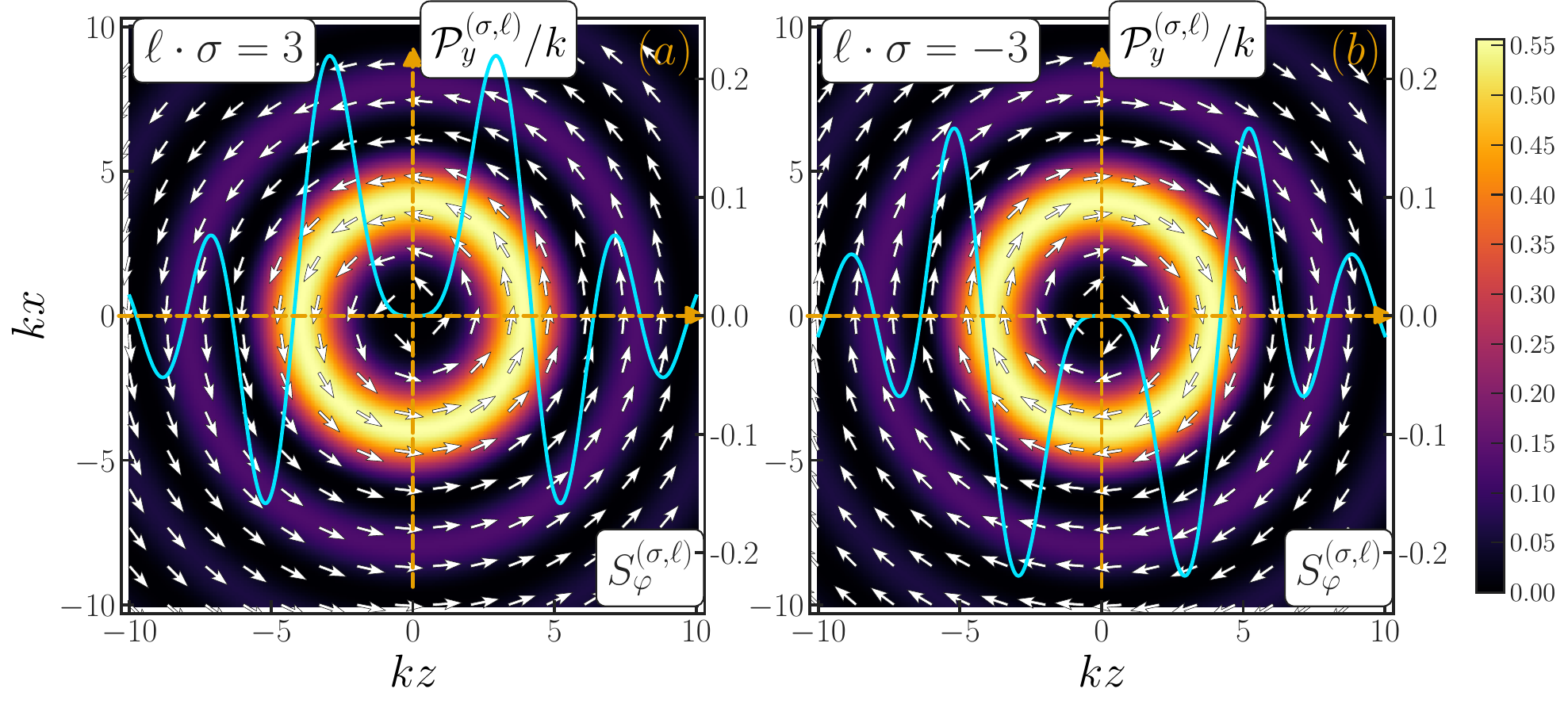} 
    \includegraphics[width = 0.98 \columnwidth]
    {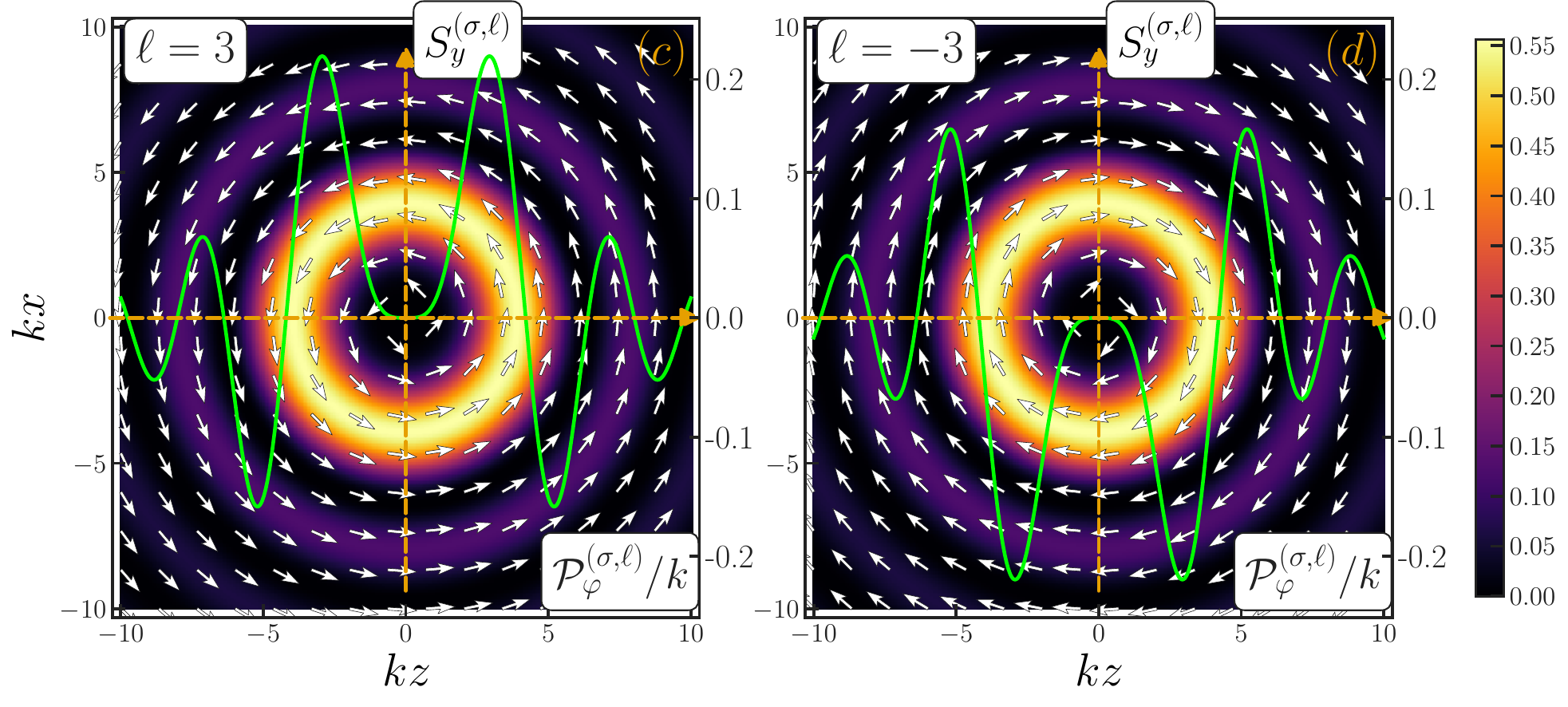}           
    \caption{Dual transverse vortex-spin-momentum locking. (a),(b) Vortex-driven spin-momentum locking of transverse Poynting momentum. The spatial distributions of the azimuthal SAM and the transverse Poynting momentum with the joint chirality $\ell\cdot\sigma=\pm 3$, respectively. (c),(d) Vortex-driven spin-momentum locking of transverse SAM. The spatial distributions of the azimuthal Poynting momentum and the transverse SAM with $\ell=\pm 3$, respectively.}
    \label{T-SVPM}
\end{figure}

Via Eq.~\eqref{eq:pps}, it is straightforward to verify that the spin momentum density can be expressed as
\begin{equation}
P^{(\sigma,\ell)}_{{\rm s},\varphi} = \frac{1}{2}\big[\na\times({S}^{(\sigma,\ell)}_y\hat{\bm{y}})\big] = \mathcal{P}^{(\sigma,\ell)}_{\varphi}-P^{(\sigma,\ell)}_{\varphi},
\label{eq:spinmomentum1}
\end{equation}
\begin{equation}
P^{(\sigma,\ell)}_{{\rm s},y} = \frac{1}{2}\big[\na\times({S}^{(\sigma,\ell)}_{\varphi}\hat{\bm{\varphi}})\big] = \mathcal{P}^{(\sigma,\ell)}_{y}.
\label{eq:spinmomentum2}
\end{equation}
The $\sigma$-independent transverse SAM generates the $\sigma$-independent azimuthal spin momentum; the $\sigma$-dependent azimuthal SAM in turn gives rise to the $\sigma$-dependent transverse spin momentum that is just the transverse Poynting momentum. With the topological charge $\ell$ fixed, since ${S}^{(\sigma,\ell)}_{\varphi}$ and  $P^{(\sigma,\ell)}_{{\rm s},y}$ depend jointly on $\ell$ and $\sigma$, the transformation $\sigma \to -\sigma$ reverses the direction of the transverse Poynting momentum, as illustrated in Figs.~\ref{T-SVPM}(a) and~\ref{T-SVPM}(b) where $\sigma \ell$ is treated as the joint chirality index.

Recently, Vernon et al.~\cite{Verno24} proposed a spin decomposition of light dual to the decomposition of momentum~\eqref{eq:pps}, in which the total SAM density is split into canonical and Poynting contributions 
\begin{equation}
\bm{S} = {\bm S}_{\rm c} + \bm{S}_{\rm p} = \bm{S}_{\rm c} + \frac{1}{2k^2}\na\times\bm{\mathcal{P}},
\end{equation}
with the canonical SAM density
\begin{equation}
\bm{S}_{\rm c}= \frac{g}{2k}{\mathrm Re}\big[E_{j}^{*}\na B_{j} - B_{j}^{*} \na E_{j}\big]. \label{Scan}
\end{equation} 

We can now position our own work within this framework as an illustration. Substituting the electromagnetic fields~(\ref{eq:spineb}) into Eq.~\eqref{Scan}, we readily obtain the canonical SAM density 
\begin{equation}
\bm{S}^{(\sigma,\ell)}_{\rm c}=\frac{\sigma}{\rho}\big[(\ell-1)J_{\ell-1}^2+2\ell J_{\ell}^{2}+(\ell+1)J_{\ell+1}^2\big]\hat{\bm{\varphi}},
\label{eq:csam}
\end{equation}
It is entirely azimuthal and exhibits a direct correlation with the canonical momentum density:
\begin{equation}
\bm{S}^{(\sigma,\ell)}_{\rm c}=\frac{\sigma}{k}\bm{P}^{(\sigma,\ell)},
\label{eq:csp}
\end{equation}

Indeed, the SAM and the Poynting momentum densities also share the same vector structure in such circularly polarized rotating-wave structured light:
\begin{equation}
\bm{S}^{(\sigma,\ell)} = \frac{\sigma}{k}\bm{\mathcal{P}}^{(\sigma,\ell)}.
\label{eq:MS}
\end{equation}
This not only preserves the multiple degrees of freedom of structured light but also restores the ``clean'' spin-momentum locking of circularly polarized plane waves. 

Likewise, the Poynting SAM density can be written in explicit component form as
\begin{equation}
S^{(\sigma,\ell)}_{{\rm p},\varphi} = \frac{1}{2k^2}\big[\na\times(\mathcal{P}^{(\sigma,\ell)}_y\hat{\bm{y}})\big] = S^{(\sigma,\ell)}_{\varphi}-S^{(\sigma,\ell)}_{{\rm c},\varphi},
\label{eq:mosp1}
\end{equation}
\begin{equation}
S^{(\sigma,\ell)}_{{\rm p},y} = \frac{1}{2k^2}\big[\na\times(\mathcal{P}^{(\sigma,\ell)}_{\varphi}\hat{\bm{\varphi}})\big] = S^{(\sigma,\ell)}_{y}.
\label{eq:mosp2}
\end{equation}
The $\sigma$-dependent transverse Poynting momentum generates a $\sigma$-dependent azimuthal SAM, while the $\sigma$-independent azimuthal  Poynting momentum yields a $\sigma$-independent transverse SAM. The transformation $\ell\to -\ell$ reverses the transverse $\sigma$-independent SAM direction, as shown in Figs.~\ref{T-SVPM}(c) and \ref{T-SVPM}(d).

Equations (\ref{eq:spinmomentum1})-(\ref{eq:spinmomentum2}) and (\ref{eq:mosp1})-(\ref{eq:mosp2}) exhibit a perfectly dual form between the SAM and Poynting momentum densities. Equations~\eqref{eq:spinmomentum2} and \eqref{eq:mosp2} together constitute a dual spin-momentum locking of the transverse SAM and transverse Poynting momentum, and both signs of these transverse quantities are subject to change with the vortex handedness. Specifically, Eq.~\eqref{eq:spinmomentum2} shows that the polarization handedness determines the sign of the transverse Poynting momentum, while Eq.~\eqref{eq:mosp2} shows that the transverse SAM is locked to the azimuthal Poynting momentum. Together, they imply a self-consistent transverse spin-momentum locking in free-space propagating waves, unlike previous observations in evanescent-field modes.

\section{Discussion and Conclusion}
To summarize, we have successfully demonstrated the circularly polarized rotating-wave structured light supports a pair of dual transverse vortex-spin-momentum locking, one in the transverse Poynting momentum and one in the transverse SAM. The transverse Poynting momentum obeys a $\mathbb{Z}_2 \times \mathbb{Z}_2$ symmetric locking structure: it is controlled independently by either the polarization handedness or the vortex handedness, yet is invariant under their simultaneous reversal. In parallel, the transverse SAM shows independence of polarization but a direct dependence on the vortex topological charge, inheriting this vortex-dependence from the Poynting momentum through the curl operation. The sign of the transverse SAM is likewise locked to the direction of the azimuthal Poynting momentum governed by the vortex handedness, which is made manifest as a transverse vortex-spin-momentum locking.

Strictly speaking, this pairing is not dual, because the transverse SAM can occur independently of the transverse Poynting momentum, as a result of their distinct polarization dependence. For instance, when the rotating-wave structured light is linearly polarized, the transverse Poynting momentum vanishes while the transverse SAM persists unchanged. However, under this configuration, a pure transverse SAM is isolated while the lateral transverse Poynting momentum is suppressed, yielding a unique lateral mechanical response that points to new possibilities for vortex-driven spin-based applications.

Overall, we achieve dual spin-momentum locking in circularly polarized rotating-wave structured light, driven by the transverse vortex via transverse spin-orbit interaction in free space. These results not only expand the understanding of vortex-controlled transverse optical linear and angular momenta in structured optical fields, but might also offer great potential for vortex-spin-controlled optical manipulation of matter.

\section*{Acknowledgments}

I sincerely acknowledge Prof. Chen for his invaluable guidance and support during this work. I also wish to express my deep gratitude to my family for their love, understanding, and support.

\end{document}